# Resonant edge-state switching in polariton topological insulators

Yiqi Zhang,[1] Yaroslav V. Kartashov[2,3,4,*], Yanpeng Zhang,[1] Lluis Torner[2] and Dmitry V. Skryabin[4,5]

[1]Key Laboratory for Physical Electronics and Devices of the Ministry of Education & Shaanxi Key Lab of Information Photonic Technique, Xi'an Jiaotong University, Xi'an 710049, China
[2]ICFO-Institut de Ciencies Fotoniques, The Barcelona Institute of Science and Technology, 08860 Castelldefels (Barcelona), Spain
[3]Institute of Spectroscopy, Russian Academy of Sciences, Troitsk, Moscow Region 108840, Russia
[4]Department of Physics, University of Bath, Bath BA2 7AY, United Kingdom
[5]ITMO University, Kronverksky Avenue 49, St. Petersburg 197101, Russia

**Abstract:** Topological insulators are unique devices supporting unidirectional edge states at their interfaces. Due to topological protection, such edge states persist in the presence of disorder and do not experience backscattering upon interaction with defects. We show that despite the topological protection and the fact that such states at the opposite edges of a ribbon carry opposite currents, there exists a physical mechanism allowing to resonantly couple topological excitations propagating at opposite edges. Such mechanism uses weak periodic temporal modulations of the system parameters and does not affect the internal symmetry and topology of the system. We study this mechanism in truncated honeycomb arrays of microcavity pillars, where topological insulation is possible for polaritons under the combined action of spin-orbit coupling and Zeeman splitting in the external magnetic field. The temporal modulation of the potential leads to a periodic switching between topological states with the same Bloch momentum, but located at the opposite edges. The switching rate is found to increase for narrower ribbon structures and for larger modulation depth, though it is changing nonmonotonically with the Bloch momentum of the input edge state. Our results provide a promising realization of a coupling device based on topologically protected states. The proposed scheme can be used in other topological photonic and condensed matter systems, including Floquet insulators and graphene.

## 1. Introduction

The physics of topological insulators attracts nowadays considerable attention across a wide range of systems [1,2]. When a topological insulator, possessing a bandgap separating two energy bands with different topological invariants (Chern numbers), interfaces with a material having a distinct topology, topologically protected in-gap states appear that are localized and propagate along the interface. Such edge states usually exhibit unidirectional propagation, with the direction of propagation controlled for example by the direction of the applied magnetic field or by swapping the order of materials across the interface. The edge states represent extremely robust topologically protected excitations that generally cannot be destroyed by perturbations with energies smaller than the energy associated with the topological gap [1,2].

Topological insulators were first studied in the context of the quantum Hall effect [1,2,3,4], while nowadays the concept has become truly interdisciplinary, thus topological insulators were proposed and experimentally demonstrated in acoustics [5,6], mechanics [7], as well as in atomic [8-11] and photonic systems [12-23]. In particular, photonic topological systems include gyromagnetic photonic crystals [12,13], semiconductor quantum wells [14], arrays of coupled resonators [15,16], metamaterial superlattices [17] and other periodic metamaterial structures [18,19], helical waveguide arrays [20–23], non-Hermitian guiding structures [24], and microcavities supporting exciton-polaritons [25-28]. Including nonlinear effects substantially enriches the behaviour of modes in topological systems [29], leading to, e.g., nonlinearity-mediated inversion of the propagation direction of the edge states [30,31], dynamical instabilities [32,33], formation of the topological edge solitons [34-36] and vortices [37], bistability [38], or nonlinear optical isolation [39].

The topological protection of edge states that makes them excellent candidates for future information processing [40-42] is simultaneously an obstacle for the realization of complex switching architectures, because immunity to local transverse perturbations of the potential, such as disorder or missing elements, implies suppressed scattering into the bulk of the insulator. Because scattering into the bulk is suppressed, topological states at the opposite edges of the standard topological insulator cannot effectively couple even if their structure is locally perturbed due to missing elements, disorder, variations in lattice periodicity, and other types of most harmful deformations that do not change with time. This poses a problem of development of novel techniques of using topological states in photonic circuits, largely relying on coupling elements, that allow to achieve a controllable transfer of topological states across the circuit. For example, a topological pump leading to mode switching between opposite edges of a topologically nontrivial structure was demonstrated in quasi-periodic one-dimensional arrays with adiabatically-varying spacing between channels [43]. Simi-larly, uniaxial strain of a honeycomb lattice can be used for the creation or destruction of states at different edges [44]. Also, utilization of transverse potential gradients parallel to the interface of the topological insulator leads to nontrivial transitions between edge states upon Bloch oscillations [45,46]. However, all these approaches require considerable modification of the spatial shape of the underlying lattice structure, and hence of the system topology.

Our main motivation here is showing switching with topological edge states that, on the one hand, would benefit from topological protection (i.e., propagation free from scattering losses and delocalization due to local inhomogeneities) without using interfaces or coupling elements that do not support edge states and therefore lead to considerable radiation emission, and that, on the other hand, would allow for complete and periodic power transfer from one strongly localized state to another strongly localized state without coupling to unprotected bulk modes. The principal aim is achieving such a coupling without modifying the very topology of the system, which at any moment of time should support the existence of well-isolated in-gap edge states. Realization of such a switching is an important step forward in the program of construction of practical topological devices.

Thus, we report a new approach allowing to resonantly couple topological edge states at the two edges of the topological insulator ribbons. The new method is based on weak periodic temporal modulations of the system parameters (e.g., depth of the underlying potential), which do not affect its spatial symmetry and topology. To illustrate the application of such approach, we consider polariton topological insulators realized as a narrow honeycomb ribbon of microcavity pillars [25,35]. Such microcavity arrays can be readily fabricated experimentally and they were already used for the experimental observation of nontopological polaritonic edge states [47,48]. Existence of topological edge states in such spinor systems [25,35] relies on the combination of polarization-dependent tunneling between neighboring pillars, which is formally analogous to spin-orbit coupling effect from condensed-matter physics [49-52], and Zeeman splitting by a magnetic field through the excitonic component of the polariton condensate.

Time-dependent modulation of optical microcavities can be realized, e.g., using electro-optic [53], acoustic [54] and free-carrier based [55] modulation techniques. Refs. [54,56,57] discuss the modulation techniques in the specific context of the polaritonic microcavities. Photonic variants of topological insulators realized as arrays of helical waveguides [20] also allow modulations of their parameters in the direction of light propagation through direct variation of the refractive index contrast in each waveguide (controlled by writing speed). Paraxial wave equation governing light evolution in such systems is formally identical to Schrödinger equation describing evolution of quantum-mechanical wavefunction in time-modulated potential.

Coupling between the edge states that we are reporting shows a strong dependence on the modulation frequency, the value of the momentum of the input edge state, and the nonlinearity strength. The physics underlying the proposed coupling mechanism is similar to the so-called stimulated mode conversion [58-62], also known as Rabi oscillations [63] in arrays of photonic waveguides [64,65].

## 2. The model and spectrum of the system

Evolution of the spinor wavefunction $\Psi=(\psi_+,\psi_-)^T$ describing polariton condensate in a narrow ribbon of the space and time periodic potential is governed by the system of dimensionless coupled Schrödinger equations [25,35]:

$$i\frac{\partial \psi_\pm}{\partial t} = -\frac{1}{2}\left(\frac{\partial^2}{\partial x^2}+\frac{\partial^2}{\partial y^2}\right)\psi_\pm + \beta\left(\frac{\partial}{\partial x}\mp i\frac{\partial}{\partial y}\right)^2 \psi_\mp + \quad (1)$$
$$\mathcal{R}(x,y)[1+\mu\sin(\omega t)]\psi_\pm \pm \Omega\psi_\pm + g(|\psi_\pm|^2 + \sigma|\psi_\mp|^2)\psi_\pm.$$

Here $\psi_\pm$ are the complex amplitudes of the spin-positive and spin-negative polaritons; parameter $\beta$ accounts for the spin-orbit coupling [25,35,38]; the term $\sim \Omega$ is the Zeeman energy splitting in the external magnetic field. Nonlinear terms proportional to the densities of the spin-positive and spin-negative polaritons originate from the repulsion of the same spin polaritons, while $\sigma=-0.05$ corresponds to the weak cross-spin attraction [66]. $g$ is the nonlinear coefficient, which can be scaled to 1, but is retained for the sake of future convenience. The honeycomb potential landscape is modelled by $\mathcal{R}(x,y) = -p\sum_{n,m}\mathcal{Q}(x-x_n,y-y_m)$, where all individual potential wells have depths $p$ and Gaussian profiles $\mathcal{Q}=\exp[-(x^2+y^2)/d^2]$, while separation between neighboring wells equals to $a$. Eqs. (1) are known to accurately describe the interactions of polaritons with various inhomogeneities in the cavity, whose presence is approximated by the inclusion of proper potential terms (see reviews [56] and Refs. [25,26,31]). Examples of the honeycomb ribbons with various widths and zigzag-zigzag interfaces are presented in the insets in Fig. 1. These structures are periodic in the $y$ direction: $\mathcal{R}(x,y)=\mathcal{R}(x,y+T)$, where period $T=3^{1/2}a$. Both spatial coordinates in Eq. (1) are normalized to the characteristic distance $L$, all energy parameters (such as potential depth and Zeeman splitting) are normalized to the characteristic energy $\epsilon_0 = \hbar^2/mL^2$, where $m$ is the effective polariton mass, while evolution time is normalized to $\hbar\epsilon_0^{-1}$, see [35] for further details. At this point we do not include into the model losses that are characteristic of polaritons – we address their impact in section 3.

Main novelty of the present model consists in small temporal periodic modulation of the lattice depth with frequency $\omega$ and amplitude $\mu\ll 1$. As mentioned above this can be realized in practice by different methods described in [53-57]. Importantly, such weak modulation does not change internal symmetry and topology of the structure, in contrast to strong variations of the underlying structure used in [43-46]. As shown below, this periodic modulation, despite being weak, is capable of inducing strong coupling between counter-propagating topological states on the two edges of a ribbon. To illustrate our findings we use $\beta=0.15$, $\Omega=0.5$, $a=1.4$, $d=0.5$, $p=8$, but similar results have been obtained for other parameters ensuring opening of the topological gap.

The necessary ingredient for the existence of topological edge states in honeycomb polariton insulators is simultaneous presence of spin-orbit coupling and Zeeman splitting. These two effects acting together break time-reversal invariance in the Eq. (1) that, according to seminal work [3] and to our simulations, gives rise to the topological gap in the vicinity of Dirac points, with the width of the gap progressively increasing with increase of the spin-orbit coupling strength $\beta$. To illustrate the impact of these effects on the spectrum of modes of the ribbon, we first consider the energy spectrum of our system without temporal modulation ($\mu=0$) and neglecting nonlinearity. Linear modes in this limit are the Bloch states periodic in $y$ and localized along $x$. They are given by $\psi_\pm(x,y,t)=u_\pm(x,y)\exp(iky+i\epsilon t)$, where $u_\pm(x,y)=u_\pm(x,y+T)$ and $u_\pm(x\to\pm\infty,y)=0$, $k$ is the Bloch momentum, and $\epsilon$ is the energy. The latter is a periodic function of $k$ with a period $K=2\pi/T$. The resulting eigenvalue problem is

$$\epsilon u_\pm = (1/2)[\partial^2/\partial x^2 + (\partial/\partial y + ik)^2]u_\pm - \mathcal{R}(x,y)u_\pm - \beta[\partial/\partial x \mp i(\partial/\partial y + ik)]^2 u_\mp \mp \Omega u_\pm. \quad (2)$$

Eqs. (2) were solved numerically using plane-wave expansion method for different widths of the ribbons. Figures 1(a)-1(c) show $\epsilon(k)$ spectra for ribbons, whose unit cells contain $22$, $14$, or $6$ microcavity pillars (insets in Fig. 1 show five such unit cells in the vertical direction, i.e. five $y$-periods of the structure). It should be mentioned that selection of the appropriate width of the ribbon is essential, since strength of the edge to edge coupling strongly depends on the spatial overlap of the edge states within the ribbon. Black curves in Fig. 1 correspond to the bulk states, while red and blue curves are associated with topological edge states. We show here only the topological gap that opens between first and second bands, which were touching in two Dirac points at $k = K/3$ and $k = 2K/3$ [their remnants are resolvable in Fig. 1(a) obtained for wider ribbon]. Note, that topological edge states can be encountered in other gaps, which are not shown here.

One can see that with decrease of the width of the ribbon the number of bulk modes in a given group of bands decreases, while there is always two in-gap topological states, suggesting that topological effects persist even when ribbons are narrow. The curve with $\partial\epsilon/\partial k > 0$ corresponds to edge states at the right edge moving in negative $y$ direction, while the curve with $\partial\epsilon/\partial k < 0$ corresponds to states at the left edge moving in the positive $y$ direction. The examples of edge states supported by narrowest ribbon at $k = 0.55\,\mathrm{K}$ and corresponding to the blue and red dots in Fig. 1(c), are shown in Fig. 2. Notice that these states are well-localized at corresponding boundaries, but at the same time they do overlap by their tails. For both edge states the spin-negative component $\psi_-$ has larger amplitude than the spin-positive component $\psi_+$, which is controlled by the magnetic field direction (sign of $\Omega$).

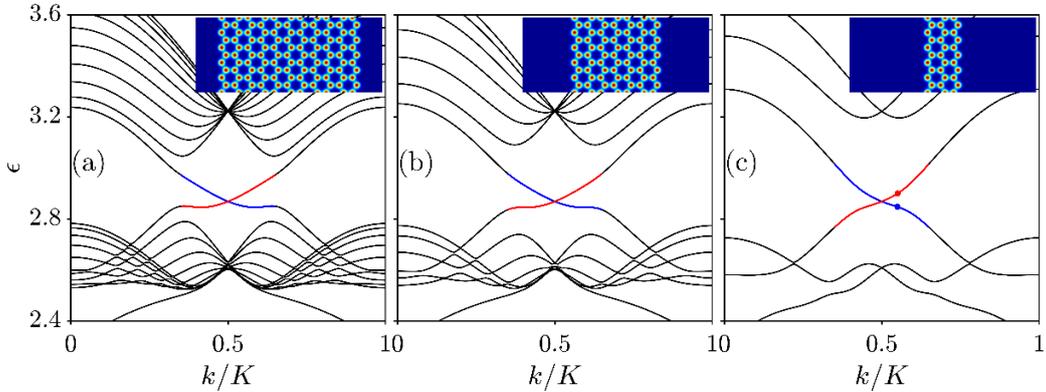

Fig. 1. (Color online) Energy-momentum diagrams $\epsilon(k)$ obtained for ribbons with zigzag edges and $22$ (a), $14$ (b), and $6$ (c) microcavity pillars within unit cell. Black curves correspond to modes residing in the bulk of the ribbon, while color lines indicate edge states. Insets show corresponding honeycomb ribbons with zigzag-zigzag edges. Dots in (c) correspond to modes at $k = 0.55\,\mathrm{K}$ shown in Fig. 2.

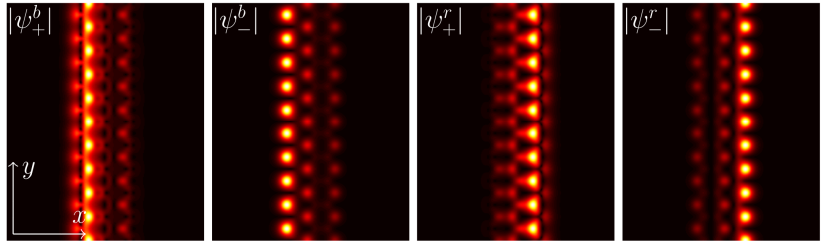

Fig. 2. (Color online) Profiles of the edge states corresponding to the blue (left two panels) and red (right two panels) dots in Fig. 1(c) at $k = 0.55\,\mathrm{K}$. Modes are shown within $x \in [-10, +10]$ and $y \in [-5T, 5T]$ windows.

## 3. Topological mode switching

When edge states obtained from (2) are used as the input of the evolution Eqs. (1), their profiles do not change with time in the absence of temporal modulation of the potential. Different edge states remain decoupled (norm per $y$ period carried by each edge state does not change with time $t$), which is evident from the absence of any mode hybridization around the point where their dispersions cross. Even if two edge states from different branches with the same momentum $k$ are excited simultaneously, one will observe only small beatings of peak amplitude due to different energies, $\epsilon_r$ and $\epsilon_b$, of linearly interfering states. Inclusion of localized defects into underlying structure also does not lead to efficient coupling of states at the opposite edges. This was verified for the narrowest ribbon that we constructed [see inset in Fig. 1(c)] with a localized defect in the form of a missing pillar on the either left or right edges. Although such a defect causes inevitable distortion of the extended edge state launched into system, the weight of the edge state on the opposite edge remains negligible (although not identically zero) over considerable time intervals that indicates on practical absence of coupling.

Inclusion of temporal modulation of the potential qualitatively changes this picture and leads to coupling between topological modes, provided that the frequency of modulation $\omega$ is properly selected. To explain such effect, we recall that the in-gap edge states are well-separated in energy from the bulk modes, so that for weak potential modulations $\mu \ll 1$ one can employ coupled-mode theory limiting consideration only to two edge states and disregarding their interaction with the in-band ones. Therefore, by assuming that the input wavefunction includes only two edge states from the opposite edges (blue and red branches), one can write the solution of Eq. (1) at $\mu \neq 0$ in the form:

$$\boldsymbol{\Psi} = c_\mathrm{r}(t) \mathbf{U}_\mathrm{r} \exp(i\epsilon_\mathrm{r} t + iky) + c_\mathrm{b}(t) \mathbf{U}_\mathrm{b} \exp(i\epsilon_\mathrm{b} t + iky), \quad (3)$$

where $c_\mathrm{r,b}(t)$ are slowly varying complex amplitudes of the modes from different branches (for convenience, subscripts are selected in accordance with colors used in Fig. 1 to denote different branches of the edge states), and $\mathbf{U}_\mathrm{r,b} = (u_+^\mathrm{r,b}, u_-^\mathrm{r,b})^\mathrm{T}$ are spinors given by Eq. (2) and describing Bloch mode shapes with selected momentum $k$. They satisfy the conditions $\langle \mathbf{U}_\mathrm{r}, \mathbf{U}_\mathrm{r} \rangle = \langle \mathbf{U}_\mathrm{b}, \mathbf{U}_\mathrm{b} \rangle = 1$ and $\langle \mathbf{U}_\mathrm{r}, \mathbf{U}_\mathrm{b} \rangle = 0$, where angular brackets correspond to the Hermitian inner product that is calculated over one unit cell of the ribbon taking into account $y$-periodicity of all modes. By substituting wavefunction in this form into Eqs. (1), neglecting rapidly oscillating terms, and projecting onto spinors $\mathbf{U}_\mathrm{r,b}$ one arrives at the coupled-mode equations:

$$\frac{dc_\mathrm{r}}{dt} = -\frac{\mu}{2} \kappa c_\mathrm{b} \exp(+i\delta t), \quad \frac{dc_\mathrm{b}}{dt} = +\frac{\mu}{2} \kappa^* c_\mathrm{r} \exp(-i\delta t), \quad (4)$$

where complex coupling constant is given by

$$\kappa = \langle \mathbf{U}_\mathrm{r}, \mathcal{R} \mathbf{U}_\mathrm{b} \rangle = \int_{-\infty}^{\infty} dx \int_0^T dy \, \mathbf{U}_\mathrm{r}^\dagger \mathcal{R} \mathbf{U}_\mathrm{b}. \quad (5)$$

Here $\dagger$ stands for Hermitian conjugation, while the function $\mathcal{R}$, defined after Eq. (1), describes the potential landscape. $\delta = \omega - (\epsilon_\mathrm{r} - \epsilon_\mathrm{b})$ is the detuning between the modulation frequency and off-set of the energies of the edge states. These equations describe resonant coupling process for topological edge states with energies $\epsilon_\mathrm{r}$ and $\epsilon_\mathrm{b}$. Here both energies are taken for the same fixed Bloch momentum $k$. Coupling time corresponding to the complete state switching from one edge to the other is expected to be shortest for the exact resonance, $\omega = \epsilon_\mathrm{r} - \epsilon_\mathrm{b}$. The general solution of Eqs. (4) can be written as

$$\begin{aligned} c_\mathrm{r} &= [c_1 \exp(+i\delta_\mathrm{eff} t / 2) + c_2 \exp(-i\delta_\mathrm{eff} t / 2)] \exp(+i\delta t / 2), \\ c_\mathrm{b} &= [c_3 \exp(+i\delta_\mathrm{eff} t / 2) + c_4 \exp(-i\delta_\mathrm{eff} t / 2)] \exp(-i\delta t / 2), \end{aligned} \quad (6)$$

where $\delta_\mathrm{eff} = (\delta^2 + \mu^2 |\kappa|^2)^{1/2}$, while the constant coefficients $c_{1-4}$ should be determined from the initial conditions at $t = 0$ taking into account the following relations $c_1(\delta + \delta_\mathrm{eff}) = i\mu\kappa c_3$, $c_2(\delta - \delta_\mathrm{eff}) = i\mu\kappa c_4$, and $c_3(\delta - \delta_\mathrm{eff}) = i\mu\kappa^* c_1$, $c_4(\delta + \delta_\mathrm{eff}) = i\mu\kappa^* c_2$. Equations (6) show that the parameter $\delta_\mathrm{eff}$ determines characteristic temporal scale of variation of modal amplitudes $t_\mathrm{s} = \pi / \delta_\mathrm{eff}$, which has the physical meaning of the edge to edge switching time. In resonance, at $\delta = 0$, the switching time is inversely proportional to the modulus of the complex coupling constant $|\kappa|$ and the modulation depth $\mu$ of potential. This means that fastest switching is observed in narrow ribbons, where overlap integral (5) calculated for wavefunctions describing edge states at the opposite edges is maximal, hence coupling constant $|\kappa|$ is largest. Notice that even though edge states at the opposite edges slightly overlap spatially, especially in narrow ribbons, they remain orthogonal to each other, i.e. $\langle \mathbf{U}_\mathrm{r}, \mathbf{U}_\mathrm{b} \rangle = 0$, even though both amplitudes $c_\mathrm{r,b}$ defining weights of edge states (i.e. populations of edge potential wells) are nonzero. Although the coupling constant is always complex, in contrast to coupling constant describing mode conversion in modulated optical guiding structures, it enters Eqs. (6) in such a way that the switching dynamics remains conservative ($|c_\mathrm{r,b}|$ oscillate periodically).

To illustrate the above predictions, we solved Eqs. (1) in the linear approximation with the input conditions (3), where initial modal amplitudes were $c_\mathrm{r}|_{t=0} = (1-\alpha)^{1/2}$ and $c_\mathrm{b}|_{t=0} = \alpha^{1/2}$, where $\alpha = 0.01$. Thus, the edge state on the right boundary initially had much higher amplitude than its counterpart on the left boundary. We used narrowest ribbon depicted in the inset of Fig. 1(c). The input edge states were taken at $k = 0.55$ K – they correspond to red and blue dots. Instantaneous modal amplitudes can be calculated from projections of the spinor $\boldsymbol{\Psi}$ on the edge states: $c_\mathrm{r,b}(t) \sim \langle \mathbf{U}_\mathrm{r,b} e^{iky}, \boldsymbol{\Psi} \rangle$. It is also convenient to introduce modal weights $\nu_\mathrm{r,b} = |c_\mathrm{r,b}|^2$. Typical dependence of the modal weights $\nu_\mathrm{r,b}$ on time in the case of exact resonance ($\delta = 0$) is shown in Fig. 3(a). The fast small-amplitude oscillations in $\nu_\mathrm{r,b}$ are due to the temporal modulation of the potential and have the same characteristic scale. Coupling between two modes stimulated by the temporal modulation of the underlying potential occurs, however, on much larger temporal scales. The power is first nearly completely transferred into the edge state on the left boundary with maximal weight of this state achieved at $t = t_2$, but then the process reverses and power starts to flow back into the edge state on the right boundary. At $t = t_4$ the initial ratio of modal weights is reproduced and power again concentrates on the right boundary. The process is therefore fully periodic and continues over indefinitely long time intervals. Figure 3 also illustrates power conservation law, since at any moment of time the condition $\nu_\mathrm{r}(t) + \nu_\mathrm{b}(t) = 1$ is satisfied. This conservation law can also be derived directly from coupled-mode Eqs. (4). The distributions of modulus of both spinor components, illustrating switching process, are shown in Fig. 4 for four representative moments of time marked in Fig. 3(a). Switching between opposite edges is most obvious from distributions of dominating $|\psi_-|$ component at $t = t_2$ and $t = t_4$. Notice that by stopping temporal modulation of potential at a desired moment of time, one can control balance of the edge state powers, because without the modulation it does not change in the linear system.

Since the very existence of topological edge states is not connected to the presence of losses and since the resonant switching studied here is purely linear effect, the inclusion of losses, which are characteristic of polaritons, does not affect qualitatively the switching dynamics. To illustrate this point, we added loss terms $-i\alpha\boldsymbol{\Psi}$, with $\alpha = 0.005$, into Eq. (1) and solved it with the same values of the parameters $k, \mu, \delta$ and the same input conditions that were used to generate Fig. 4. The $|\psi_-|$ distributions in the same moments of time $t_1 - t_4$ are shown in Fig. 5 and they can be compared with the distributions shown in Fig. 4 obtained without losses. Except for the obvious exponential decrease of the amplitude caused by losses, the patterns remain unchanged. Modal weights obtained in the dissipative case and multiplied by $\exp(2\alpha t)$ practically coincide with the weights shown in Fig. 3(a). The inclusion of losses may affect results only when nonlinearity is taken into account, but then the external pump that is routinely used in polariton condensates may be employed to compensate the losses.

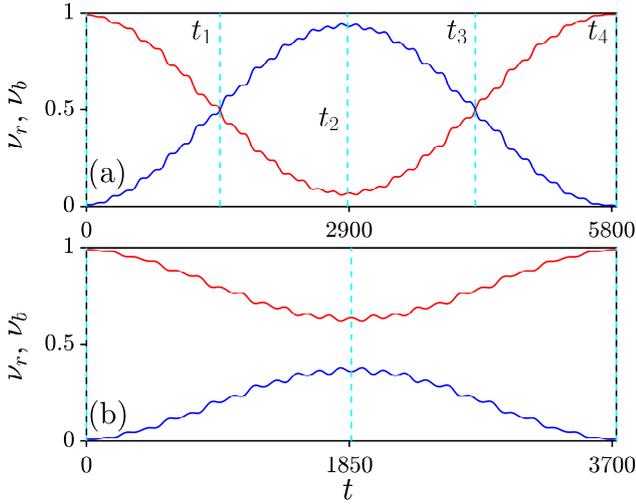

Fig. 3. (Color online) Evolution of modal weights $\nu_{r,b}$ of edge states at $k = 0.55K$ in modulated potential. (a) Resonant case $\delta = 0$. (b) Non-resonant case with $\delta = 0.001$. In both cases $\mu = 0.15$. Red and blue curves represent the weights of the edge states on the right and left edges, respectively.

Having established the possibility of resonant switching of topological edge states, we now address its potential implementation in actual samples. We consider an array formed by pillars, whose contribution to the potential landscape is described by the Gaussian potentials of $1\ \mu m$ full width, with center-to-center separation of $1.4\ \mu m$ (we assume here the characteristic length of $L=1\ \mu m$). The effective polariton mass $m \sim 10^{-34}$ kg and energy $\epsilon_0 \sim 0.7$ meV allow to get an estimate of $5.6$ meV for the potential depth, which corresponds to $p = 8$. We assume an experimentally achievable Zeeman splitting of $0.35$ meV, corresponding to $\Omega = 0.5$ [67], and a moderate contribution $0.1$ meV coming from the spin-orbit-coupling parameter $\beta = 0.15$ [50,68]. A characteristic time scale $0.9$ ps implies that resonant coupling of edge states at $k = 0.55\,K$ occurs for modulation periods $2\pi/(\epsilon_r - \epsilon_b)$ of about $250$ ps. For other momentum values this period can be smaller by a factor of $10$. Although such fast modulation is challenging, it is principally compatible with some of the techniques reported in [53-57], especially those using ultrafast pulse sequences. Observation of switching then requires a sufficiently long and narrow ribbon, where one of the edges is to be excited using a broad beam with proper momentum. Since resonant switching is a linear process and it occurs over the entire length of the ribbon, it should be observed for a large class of profiles of the input excitation. Typical length of the ribbon necessary for observation of conversion can be estimated as a distance along the edge that broad, but localized excitation traverses over typical conversion time. Taking representative value of $t_s = 2000$ for $\mu = 0.2$ at exact resonance $\delta = 0$ for $k = 0.55\,K$ we estimate this distance as $\delta y \approx 600$ [using group velocity $\partial \epsilon/\partial k$ from Fig. 1(c)] that corresponds to $600\ \mu m$ (or around $250$ periods of the structure). Example of such extended structure was fabricated in [47].

Coupling between edge states occurs not only for modulation at resonant frequency $\omega = \epsilon_r - \epsilon_b$, but also for slightly non-resonant modulation, i.e. for nonzero detuning $\delta$ from exact resonance. Representative dynamics of modal weights for $\delta = 0.001$ is shown in Fig. 3(b). On the one hand, one finds that switching is now incomplete, i.e. maximal switching efficiency for this small detuning does not reach $50\%$. On the other hand, the edge-to-edge switching time, defined as the moment, where the amplitude of the state on the left boundary reaches maximum, decreases in comparison with switching time achieved for the exactly resonant modulation, in agreement with the above theory.

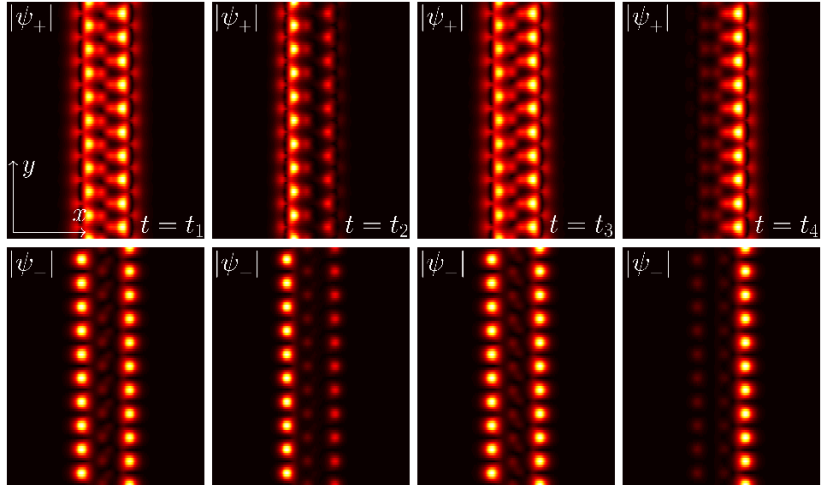

Fig. 4. (Color online) $|\psi_\pm|$ distributions showing resonant switching process at $k = 0.55\,K$, $\mu = 0.15$, and $\delta = 0$. Input states correspond to red and blue dots in Fig. 1(c). Distributions are shown within $x \in [-10, +10]$ and $y \in [-5T, 5T]$ windows.

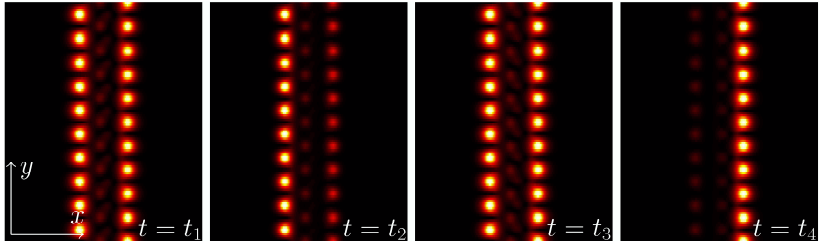

Fig. 5. (Color online) $|\psi_-|$ distributions for the same input conditions and parameters as in Fig. 4, but in the presence of losses $\alpha = 0.005$.

Dependence of the edge-to-edge switching time vs detuning $\delta$ was obtained numerically by solving original Eq. (1) with the input conditions (3) for $k = 0.55\,\mathrm{K}$. The dependence $t_c(\delta)$ obtained for the linear case $(g = 0)$ is presented in Fig. 6(a). Figure 7(a) illustrates corresponding maximal switching efficiency, i.e. maximal weight of the edge state on the left boundary $\nu_b^{\max}$ as a function of $\delta$. Switching time rapidly decreases with increase of the frequency detuning, but this is also accompanied by rapid decrease of the switching efficiency [Fig. 7(a)]. The process is therefore strongly selective, since the width of resonance in $\delta$ is of the order of $0.01$ that is substantially smaller than the $\epsilon$-width of the topological gap in Fig. 1. Note, that in the frames of the original model Eq. (1) maximal switching efficiency is achieved for very small negative detuning $\delta = -5\times 10^{-4}$, which can be attributed to the impact of the higher-order oscillating terms that were neglected in the above coupled-mode theory. Resonance curves are also slightly asymmetric, even in the linear case.

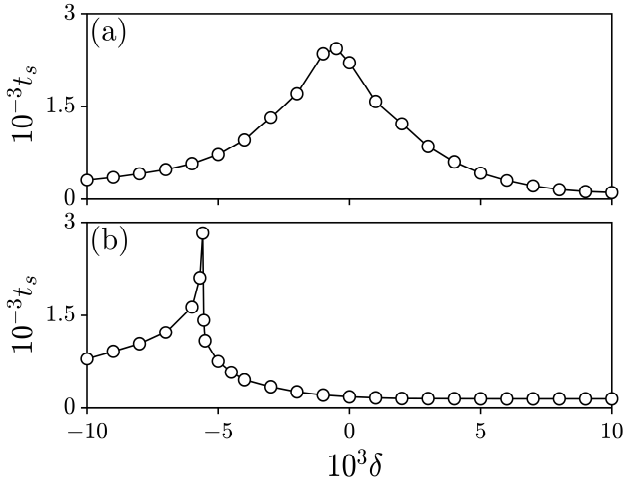

Fig. 6. Edge state switching time versus frequency detuning $\delta$. (a) Linear case. (b) Nonlinear case with $\sigma = -0.05$ and $g = 0.03$. In both cases $k = 0.55\,\mathrm{K}$ and $\mu = 0.15$.

Inclusion of even very weak nonlinearity with $g = 0.03$ for the same input makes resonance curves strongly asymmetric, see Figs. 6(b) and 7(b) for dependencies $t_s(\delta)$ and $\nu_b^{\max}(\delta)$, respectively. The resonance in terms of switching time becomes much narrower and shifts to the left [Fig. 6(a)]. The curve shows tendency to a very fast growth (nearly divergence) of switching time in narrow detuning interval. Switching efficiency in the same interval of detuning changes almost linearly and then suddenly drops [Fig. 7(b)], remaining substantially lower than peak switching efficiency achieved in the linear medium.

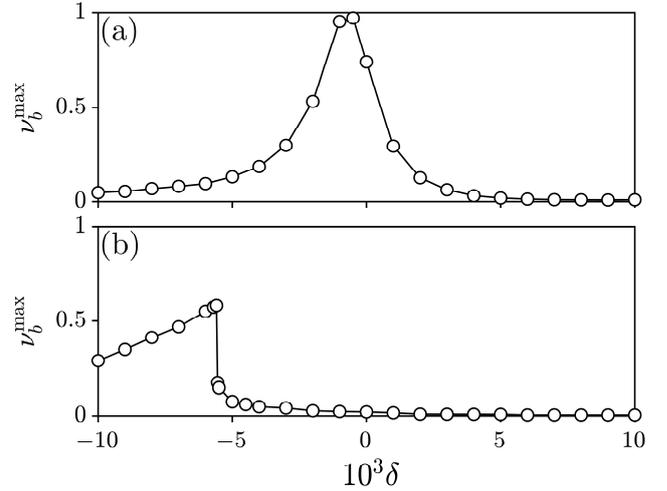

Fig. 7. Edge state switching efficiency versus frequency detuning $\delta$. (a) Linear case. (b) Nonlinear case with $\sigma = -0.05$ and $g = 0.03$. In both cases $k = 0.55\,\mathrm{K}$ and $\mu = 0.15$.

## 4. Momentum-dependent coupling

Among the distinctive properties of the edge states in topological insulators is that their localization near the edge strongly depends on the transverse momentum $k$. Generally, if $k$ is close to the crossing point of the red and blue branches in Fig. 1(c) that is located approximately in the middle of the topological gap, the localization of the edge states is strongest. However, when $k$ approaches $\mathrm{K}/3$ or $2\mathrm{K}/3$ the eigenvalue of the edge state approaches the band of the quasi-continuum and the edge state extends into the depth of the ribbon. Intuitively, one would expect monotonous decrease of switching time when Bloch momentum $k$ changes between $\mathrm{K}/2$ and $2\mathrm{K}/3$ due to growing overlap of tails of the edge states in the middle of the ribbon. However, direct simulations of Eq. (1) show rather unexpected nonmonotonic behavior of the switching time with $k$, as illustrated in Fig. 8(b) with open circles. To obtain this dependence for each value of $k$ we used resonant modulation at the frequency $\omega = \epsilon_r - \epsilon_b$. We found that the switching time diverges around degenerate point $k = \mathrm{K}/2$, where two edge states acquire identical energies and that it has a pronounced maximum close to $k \approx 0.573\,\mathrm{K}$. To explain the latter we calculated the coupling constant (5) as a function of momentum $k$ [Fig. 8(a)]. Predictably we found that the coupling is zero at the point of degeneracy of the two edge states. However, we have also found that instead of the expected monotonous growth with $k$, the modulus of the coupling constant vanishes again close to $k \approx 0.573\,\mathrm{K}$ (since $\kappa$ is complex this means that both real and imaginary parts of $\kappa$ become zero). Vanishing of coupling constant for this intermediate value of momentum is exclusively the result of specific shapes of the corresponding spinors $\mathbf{U}_{r,b}$ and potential $\mathcal{R}$: the unexpected fact is that both real and imaginary parts of $\kappa$ can become zero simultaneously. Switching time calculated from the expression $t_s = \pi / \delta_{\mathrm{eff}}$ is shown in Fig. 8(b) with the red curve. Notice excellent agreement between the coupled-mode theory and results of direct simulations. Although the theory predicts divergence of the switching time at $k \approx 0.573\,\mathrm{K}$, the modeling of Eqs. (1) shows that in reality $t_s$ stays finite. We anticipate that this is due to the contribution from the fast oscillating terms that were omitted upon derivation of Eqs. (4) and which become important exactly when $\kappa \to 0$.

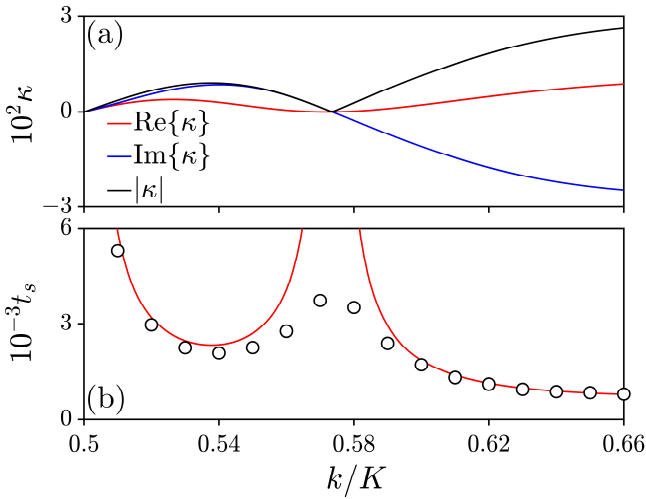

Fig. 8. (Color online) (a) Coupling coefficient $\kappa$ versus $k$. (b) Switching time $t_s$ versus $k$. Red line – prediction of coupled-mode theory, open circles – results of direct simulations of Eq. (1). In all cases $\mu = 0.15$.

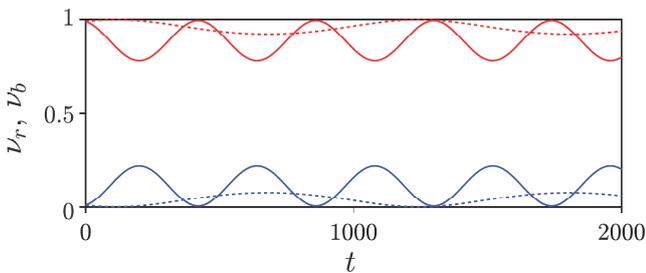

Fig. 9. (Color online) Evolution of modal weights of states at the opposite edges without temporal modulation of the potential, but in the presence of nonlinearity $g = 0.03$ at $k = 0.50\,\mathrm{K}$ (solid lines) and $k = 0.51\,\mathrm{K}$ (dashed lines).

Suppression of coupling of the edge states around $k \approx 0.573\,\mathrm{K}$ is a classic example of the situation when coupling is prohibited by symmetry reasons [edge state profiles are such that the integral (5) vanishes]. Since the perturbed potential directly enters Eq. (5), the selection of the particular type of perturbation is crucial because not all time-dependent perturbations can couple edge states, i.e. not all of them lead to nonzero $\kappa$ values. Thus, we verified that a small horizontal or vertical shaking (displacement) of the structure does not lead to coupling of states at opposite edges, because the effective contribution to the potential introduced by such shaking is antisymmetric in $x$ or $y$ and leads to a vanishing coupling constant.

It should be also mentioned that close to the point $k = 0.5\,\mathrm{K}$ in Fig. 8(b), where energies of states at the opposite edges of the topological insulator coincide and where switching between them due to the temporal modulation of the potential is practically inhibited, nonlinearity may still lead to a weak coupling even at $\mu = 0$, i.e. without modulation of the potential. Figure 9 illustrates the dynamics of the modal weights in this case. The efficiency of such a purely nonlinear coupling is relatively low and it rapidly decreases with increase of the momentum $k$ (compare the solid lines for $k = 0.50\,\mathrm{K}$ and the dashed lines for $k = 0.51\,\mathrm{K}$).

Finally, we address the impact of the modulation amplitude $\mu$ of potential on switching time $t_s$. In accordance with predictions of the coupled-mode theory at exact resonance ($\delta = 0$) the switching time decreases as $t_s \sim 1/\mu$ [see Fig. 10(a)]. This allows to substantially speed-up the switching process, however $\mu$ cannot be taken too large, because large temporal modulations of the potential lead to strong radiation into the quasi-bulk states.

While all results presented above were obtained for narrow ribbons, the phenomenon of course exist in wide structures. Naturally, because overlap of modal fields of edge states will decrease with increase of the width of the ribbon leading to reduction of coupling constant (5), one should expect notable increase of the switching time. The latter is shown in Fig. 10(b) as a function of the number of micropillars $n$ in one unit cell of the ribbon. For large $n$ (wide ribbons) the dependence $t_s(n)$ is exponential. Thus, most optimal conversion implies narrow ribbons, but we stress that even in narrowest structures that we considered [inset in Fig. 1(c)], where inclusion of defects did not result in efficient mode coupling, the edge states at the opposite edges were strongly localized [Fig. 4] and overlapped only through small tails.

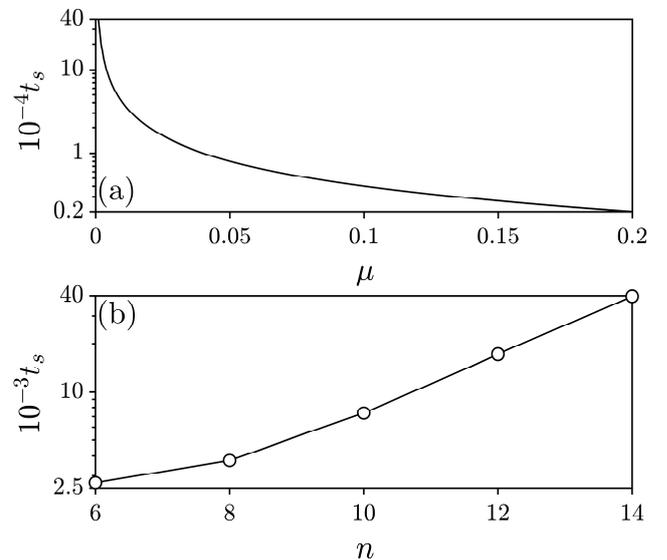

Fig. 10. (a) Switching time versus $\mu$ in the ribbon with six micropillars in the unit cell. (b) Switching time versus number of micropillars $n$ in the unit cell at $\mu = 0.15$. In all cases $k = 0.55\,\mathrm{K}$. Notice logarithmic scales in all vertical axes.

## 5. Summary

Summarizing, we put forward a new technique that can couple topologically protected edge states existing at the opposite edges of a topological insulator ribbon. The technique is based on the weak temporal modulations of the underlying potential with frequencies matching closely the energy difference between the two edge states for a given momentum. We found that the transverse momentum strongly impacts the coupling efficiency between the edge states, which thus can be inhibited for certain momentum values. Our results suggest promising possibilities for the implementation of switching devices based on topologically protected states and they motivate experiments about the controllable transformation of edge states in several physical settings where topological effects are under current intense investigation. In particular, they provide insight into the development of topological lasers that are insensitive to disorder [69,70], whose operation regimes may be controlled by means of resonant mode coupling.

**Funding**. National Key R&D Program of China (2017YFA0303703); Natural Science Foundation of Shaanxi Province (2017JZ019); Key Scientific and Technological Innovation Team of Shaanxi Province (2014KCT-10); China Postdoctoral Science Foundation (2016M600777); National Natural Science Foundation of China (11474228); ITMO University Visiting Professorship via the Government of Russia Grant 074-U01. Y.V.K. and L.T. acknowledge support from the Severo Ochoa Excellence Programme (SEV-2015-0522), Fundacio Privada Cellex, Fundacio Privada Mir-Puig, and CERCA (Generalitat de Catalunya). Y.V.K. acknowledges funding of this study by RFBR and DFG according to the research project № 18-502-12080.